\documentclass[%
reprint,
superscriptaddress,
 aps,
prstab,
floatfix
]{revtex4-2}

\usepackage{graphicx}
  \setkeys{Gin}{width=\linewidth}
\usepackage{dcolumn}
\usepackage{bm}

\usepackage{amsmath,amsfonts,amssymb}
\usepackage{amsbsy}

\usepackage{hyperref}
\hypersetup{
    colorlinks=true,       
    linkcolor=blue,          
    citecolor=blue,        
    filecolor=blue,         
    urlcolor=blue        
}

\newcommand{\orcidicon}[1]{\href{https://orcid.org/#1}{\includegraphics[height=\fontcharht\font`\B,keepaspectratio]{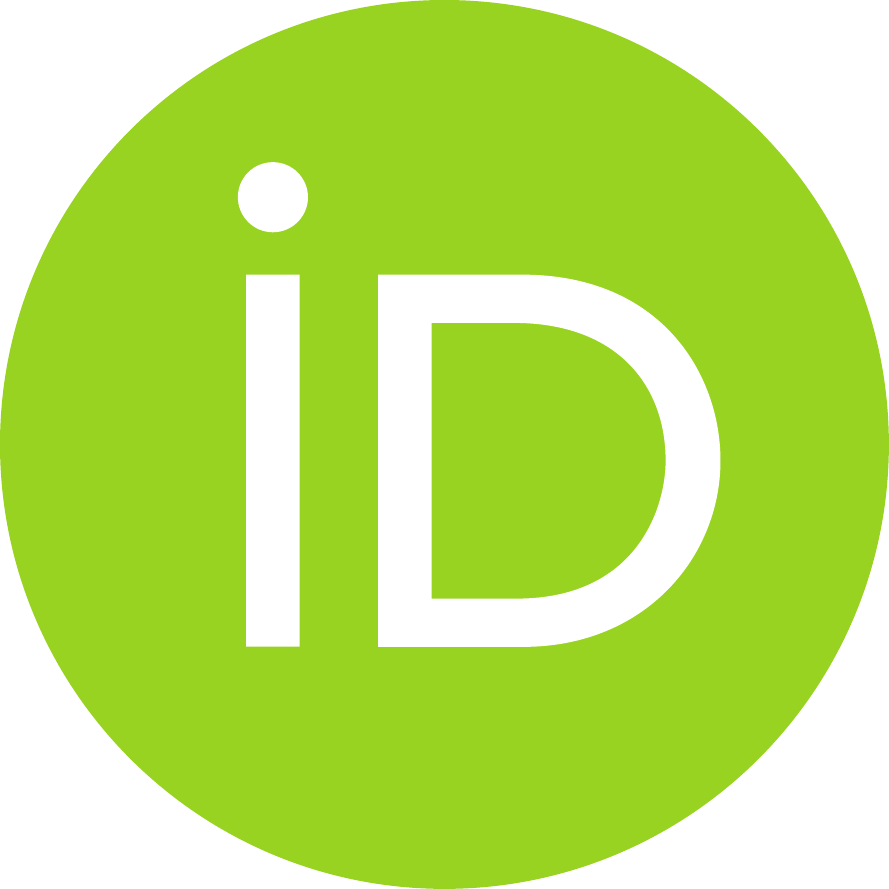}}}

\begin{document}


\title{1D finite-width graphene nanoribbon systems: alkalization and hydrogenation}

\author{Wei-Bang Li\,\orcidicon{0000-0002-8319-3316}}
\email[E-mail: ]{weibang1108@gmail.com}
\affiliation{Department of Physics, National Cheng Kung University, Tainan 70101, Taiwan}

\author{Yu-Ming Wang\,\orcidicon{0000-0001-8866-065X}}
\email[E-mail: ]{wu0h96180@gmail.com}
\affiliation{Department of Physics, National Cheng Kung University, Tainan 70101, Taiwan}

\author{Hsien-Ching Chung\,\orcidicon{0000-0001-9364-8858}}
\email[E-mail: ]{hsienching.chung@gmail.com}
\affiliation{RD Dept., Super Double Power Technology Co., Ltd., Changhua City, Changhua County, 500042, Taiwan}

\author{Ming-Fa Lin}
\email[E-mail: ]{mflin@mail.ncku.edu.tw}
\affiliation{Department of Physics, National Cheng Kung University, Tainan 70101, Taiwan}

\date{\today}

\begin{abstract}
Graphene is the first truly two-dimensional (2D) material, possessing a cone-like energy spectrum near the Fermi energy and treated as a gapless semiconductor. Its unique properties trigger researchers to find more applications of it, such as high carrier mobility at room temperature, superior thermoconductivity, high modulus and tensile strength, high transparency, and anomalous quantum Hall effect. However, the gapless feature limits the development of graphene nanoelectronics. Making one-dimensional (1D) strips of graphene (i.e., graphene nanoribbons (GNRs)) could be one of the most promising approaches to modulating the electronic and optical properties of graphene. The electronic and optical properties have been theoretically predicted and experimentally verified highly sensitive to the edge and width. The tunable electronic and optical properties further imply the possibilities of GNR application. Recently, the dangling bond problem is under consideration in the GNR system. Various passivation at the ribbon edge might change the physical properties. In this work, some passivation conditions are studied, such as alkalization and hydrogenation.
\begin{description}
\item[Keywords]
graphene nanoribbon, passivation, alkalization, hydrogenation.
\item[Usage]
This is a preprint version.
\end{description}
\end{abstract}


\maketitle


\section{Introduction}

\subsection{From graphene to strip-like graphene structures}
A new scientific field has been reached since the discovery of graphene. This material stimulates both fundamental scientists and applied technology engineers for several reasons. Graphene is the first truly two-dimensional material \cite{Science306(2004)666K.S.Novoselov, Proc.Natl.Acad.Sci.U.S.A.102(2005)10451K.S.Novoselov}, serving as an ideal platform for understanding low-dimensional physics and applications. It exhibits a cone-like energy spectrum near the Fermi energy and is treated as a gapless semiconductor. The electronic structure of graphene leads to a lot of fascinating essential properties, such as high carrier mobility at room temperature ($>200,000$ cm$^2$/Vs) \cite{Science312(2006)1191C.Berger, SolidStateCommun.146(2008)351K.I.Bolotin, Phys.Rev.Lett.100(2008)016602S.V.Morozov}, superior thermoconductivity (3,000--5,000 W/mK) \cite{NanoLett.8(2008)902A.A.Balandin, Phys.Rev.Lett.100(2008)016602S.V.Morozov}, high modulus ($\sim$1 TPa) and tensile strength ($\sim$130 GPa) \cite{Science321(2008)385C.Lee}, high transparency to incident light over a wide range of wavelength (97.7\%) \cite{Nat.Nanotechnol.5(2010)574S.Bae, Science320(2008)1308R.R.Nair}, and anomalous quantum Hall effect \cite{Nature438(2005)197K.S.Novoselov, Science315(2007)1379K.S.Novoselov, Nat.Phys.2(2006)177K.S.Novoselov, Nature438(2005)201Y.B.Zhang}.

Owing to the aforementioned superior electronic, thermal, mechanical, and optical properties, graphene is treated as high potential candidate in synthesizing future electronic and optical devices. However, the gapless feature causes a low on/off ratio in graphene-based field-effect transistors (FETs) and shrinks the development of graphene nanoelectronics. One of the most promising approaches to modulating electronic and optical properties is to make 1D strips of graphene, i.e., GNRs \cite{Phys.Chem.Chem.Phys.18(2016)7573H.C.Chung}.

There are many fabrication strategies to achieve large-scale production of GNRs, including both bottom-up and top-down schemes. From the viewpoint of geometry, graphene cutting seems to be the most simple and intuitive strategy to fabricate GNRs, and the available routes contain lithographic patterning and etching of graphene \cite{NanoLett.9(2009)2083J.W.Bai, PhysicaE40(2007)228Z.H.Chen, Phys.Rev.Lett.98(2007)206805M.Y.Han, Nat.Nanotechnol.3(2008)397L.Tapaszto}, sonochemical breaking of graphene \cite{Science319(2008)1229X.L.Li, Phys.Rev.Lett.100(2008)206803X.R.Wang, NanoRes.3(2010)16Z.S.Wu}, oxidation cutting of graphene \cite{Adv.Mater.21(2009)4487L.Ci, Chem.Mater.19(2007)4396M.J.McAllister}, and metal-catalyzed cutting of graphene \cite{NanoLett.9(2009)2600L.C.Campos, Adv.Mater.21(2009)4487L.Ci, NanoRes.1(2008)116L.Ci, NanoLett.8(2008)1912S.S.Datta, Phys.StatusSolidiB246(2009)2540F.Schaffel, NanoRes.2(2009)695F.Schaffel, NanoLett.9(2009)457N.Severin, NanoRes.3(2010)16Z.S.Wu}. A much more interesting strategy is the unzipping of carbon nanotubes (CNTs), because a CNT can be treated as a folded or zipped GNR \cite{BookLinRichQuasiparticlePropertiesLowDimensionalSystems}. The available routes for the reverse process cover chemical attack \cite{Carbon48(2010)2596F.Cataldo, Nature458(2009)872D.V.Kosynkin}, laser irradiation \cite{Nanoscale3(2011)2127P.Kumar}, plasma etching \cite{NanoRes.3(2010)387L.Jiao, Nature458(2009)877L.Jiao}, metal-catalyzed cutting \cite{NanoLett.10(2010)366A.LauraElias, Nanoscale3(2011)3876U.K.Parashar}, hydrogen treatment and annealing \cite{ACSNano5(2011)5132A.V.Talyzin}, unzipping functionalized CNTs by scanning tunneling microscope (STM) tips \cite{NanoLett.10(2010)1764M.C.Pavia}, electrical unwrapping by transmission electron microscopy (TEM) \cite{ACSNano4(2010)1362K.Kim}, intercalation and exfoliation \cite{NanoLett.9(2009)1527A.G.Cano-Marquez, ACSNano5(2011)968D.V.Kosynkin}, electrochemical unzipping \cite{J.Am.Chem.Soc.133(2011)4168D.B.Shinde}, and sonochemical unzipping \cite{Nat.Nanotechnol.5(2010)321L.Jiao, J.Am.Chem.Soc.133(2011)10394L.Xie}. Other strategies contain chemical synthesis \cite{ACSNano3(2012)2020S.Blankenburg, Nature466(2010)470J.M.Cai, J.Am.Chem.Soc.130(2008)4216X.Y.Yang, Appl.Phys.Lett.105(2014)023101Y.Zhang} and chemical vapor deposition (CVD) \cite{NanoLett.8(2008)2773J.Campos-Delgado, Nat.Nanotechnol.5(2010)727M.Sprinkle, J.Am.Chem.Soc.131(2009)11147D.C.Wei}. The former is piecewise linking of molecular precursor monomers, and the latter is much compatible with the current semiconductor industry.

The electronic and optical properties of GNRs are determined by the ribbon width and the edge structure. GNRs with armchair and zigzag edges are two characterized systems chosen for model studies. The lateral quantum confinement causes many 1D parabolic subbands, in which the electronic states are featured by the regular standing waves. Particularly, the energy gaps in armchair GNRs (AGNRs) scale inversely with the ribbon width \cite{NanoLett.6(2006)2748V.Barone, Phys.Rev.Lett.97(2006)216803Y.W.Son}. On the other hand, zigzag GNRs (ZGNRs) possess partial flat subbands near the Fermi energy with peculiar edge states localized at the ribbon edges \cite{J.Phys.Soc.Jpn.65(1996)1920M.Fujita, Phys.Rev.B54(1996)17954K.Nakada}. The former have been observed by the experimental electric conductance measurements \cite{Phys.Rev.Lett.98(2007)206805M.Y.Han, Science319(2008)1229X.L.Li} and tunneling current measurements \cite{Nat.Nanotechnol.3(2008)397L.Tapaszto} and the latter are confirmed by the STM image \cite{Phys.Rev.B73(2006)125415Y.Kobayashi, Phys.Rev.B71(2005)193406Y.Kobayashi}. As for the optical properties, the edge-dependent absorption selection rules are predicted, i.e., $|\Delta n| = odd$ for ZGNRs and $|\Delta n| = 0$ for AGNRs, where $n$ is the subband index \cite{Opt.Express19(2011)23350H.C.Chung, Phys.Rev.B76(2007)045418H.Hsu, Phys.Rev.B84(2011)085458K.Sasaki}. The essential properties will be enriched by the external fields, such as magnetic and electric fields.

A uniform static perpendicular magnetic field can accumulate the neighboring electronic states, inducing highly degenerate Landau levels (LLs) with quantized cyclotron orbits \cite{Z.Physik64(1930)629L.Landau}. The magnetic quantization can be severely suppressed by the lateral confinement. The competition between the magnetic confinement and the lateral confinement diversifies the magneto-electronic structures, containing partly dispersionless quasi-Landau levels (QLLs), 1D parabolic subbands, and partial flat subbands \cite{Phys.Rev.B73(2006)195408L.Brey, Phys.Rev.B59(1999)8271K.Wakabayashi}. Such energy subbands contribute two kinds of peaks (i.e., symmetric and asymmetric peaks) in the density of states (DOS) \cite{PhysicaE42(2010)711H.C.Chung, Carbon109(2016)883H.C.Chung}. QLLs can be survived in sufficiently wide GNRs, and each Landau wave function has a localized symmetric/antisymmetric distribution with a specific number of zero points (a quantum number $m$). The magneto-optical spectra show many symmetric and asymmetric absorption peaks. The symmetric absorption peaks come from the inter-QLL transitions and obey the magneto-optical selection rule of $|\Delta m| = 1$. However, the asymmetric absorption peaks originate from the transitions among parabolic subbands and abide by the edge-dependent selection rules \cite{Phys.Chem.Chem.Phys.18(2016)7573H.C.Chung}.

A transverse static electric field generates an extra potential energy in GNRs, making the charge carriers experience different site energies \cite{J.Phys.Soc.Jpn.80(2011)044602H.C.Chung, Philos.Mag.94(2014)1859H.C.Chung}. The electronic and optical properties are drastically changed. There exist oscillatory energy subbands, more extra band-edge states, gap modulations, semiconductor-metal transitions, and irregular standing waves. At the same time, the intensity, number, and frequency of asymmetric DOS peaks are obviously changed. The absorption peaks associated with the edge-dependent selection rules are inhibited, and more extra peaks in the optical spectra. The different potential energies in GNRs restrict the formation of Landau orbits, and QLLs would tilt, become oscillatory, or exhibit crossings and anti-crossings. Furthermore, the inter-QLL optical transitions will be seriously modified or even destroyed thoroughly \cite{Phys.Chem.Chem.Phys.18(2016)7573H.C.Chung, Phys.Chem.Chem.Phys.15(2013)868H.C.Chung}.

\subsection{Monohydrogenation and dihydrogenation}
The edges of GNRs have attracted much interest owing to their potentially strong influence on GNR electronic and magnetic properties. It is well known that the C--C bonds in GNR are well and strongly bonded. However, an edge atom would have dangling bonds, making unstable edge structures \cite{Phys.Rev.B77(2008)165427F.Cervantes-Sodi, Science323(2009)1705C.O.Girit, Phys.Rev.Lett.101(2008)115502P.Koskinen, ACSNano7(2013)198X.Zhang}, and thus each edge carbon atom is often passivated by one hydrogen atom, i.e., monohydrogenation or $sp^2$-like hydrogenation. Besides monohydrogenation, the carbon atoms at the GNR edges can also be terminated by two hydrogen atoms, i.e., dihydrogenation or $sp^3$-like hydrogenation \cite{Phys.Rev.B82(2010)165405S.Bhandary, Phys.Rev.B67(2003)092406K.Kusakabe, Phys.Rev.B79(2009)165440G.Lee, J.Phys.Chem.C115(2011)9442Y.Liu, Phys.Rev.B84(2011)075481B.Sahu, Phys.Rev.Lett.101(2008)096402T.Wassmann, Appl.Phys.Lett.96(2010)163102B.Xu}.

Recently, many researches have been reported concerning the edge modification since it can be applied to modulate the electronic properties of GNRs for various purposes. Bhandary \emph{et al.} have predicted the hydrogen passivation effect on ZGNRs by density-functional calculations and the tight-binding theory \cite{Phys.Rev.B82(2010)165405S.Bhandary}. Each edge carbon atom bonded with 2 hydrogen atoms open up an energy gap and destroys magnetism for small-width ZGNRs (less than 8 rows). However, a re-entrant magnetism accompanied by a metallic electronic structure is observed from eight rows and wider ZGNRs. The electronic structure and magnetic state are very complex for this type of termination, with $sp^3$-like edge atoms being nonmagnetic whereas the nearest neighboring atoms are metallic and magnetic. They have also evaluated the phase stability of several thicknesses of ZGNR and demonstrated that $sp^3$-like edges can be stabilized over $sp^2$-like edges at high temperatures and pressures. Zeng \emph{et al.} have studied electron transport properties of hydrogen-terminated ZGNR and oxygen-terminated ZGNR heterostructures by first-principles calculations \cite{Appl.Phys.Lett.98(2011)053101M.G.Zeng}. The heterostructure exhibited a large charge transmission gap near the Fermi energy, and rectification behaviors were observed. Zeng \emph{et al.} have investigated the electronic transport properties of edge hydrogenated ZGNR heterojunctions by using nonequilibrium Green's functions in combination with the density functional theory \cite{J.Phys.Chem.C115(2011)25072J.Zeng}. A perfect spin-filtering effect with 100\% spin polarization and a rectifying behavior with a ratio larger than $10^5$ can be reached based on the monohydrogenated and dihydrogenated ZGNR heterojunctions at finite bias voltages. Recently, researchers proposed that the composition of $sp^2$- and $sp^3$-like hydrogenations at the GNR edges can be controlled through the chemical potential of hydrogen, \cite{Appl.Phys.Lett.94(2009)122111Y.H.Lu, Phys.Rev.Lett.101(2008)096402T.Wassmann}, indicating that the monohydrogen-terminated and dihydrogen-terminated GNR heterojunction can be synthesized in the experiment.

\subsection{Alkalization}
The dangling-bond-induced instability can be solved by the passivation of other atoms, such as lithium (Li), sodium (Na), and potassium (K). Mao \emph{et al.} have studied the electronic properties of GNRs with K atoms adsorption on the GNR edges by using density-functional theory calculations \cite{Appl.Surf.Sci.280(2013)698Y.Mao}. Their simulations exhibited that the electronic structure of the AGNR is changed from semiconductor to metal due to the charge transfer from K adatoms to the edge carbon atoms. It was reported in AGNR that the potassium adatom edge-adsorbed system is always nonmagnetic and the interaction between the K adatom and AGNR is purely ionic. In ZGNR, K edge-adsorption will suppress the magnetic moment at the impurity site and its vicinity while keeping well the edge structure of ZGNR. If selectively introduce K impurities asymmetry in two sides of the ZGNR, a ferromagnetic state can be carried out, while it is antiferromagnetic in pristine ZGNRs. These results suggest that the electronic properties of GNRs can be effectively controlled by potassium edge-passivation, and that the alkali-metal edge-passivated GNRs can serve as potential materials in nanoelectronics and spintronics.

\medskip

Up to now, there have been a lot of theoretical studies on pristine GNRs, even with/without hydrogen atom and alkali metal atom passivation. The important electronic properties are sensitive to the change in edge structure; also, the energy gaps are predicted to be inversely proportional to the width of nanoribbons. The first-principles calculations within the density functional theory are useful in fully understanding the essential properties of pristine/H-passivated/alkali metal-passivated GNRs. How to induce the drastic changes by the hydrogen and alkali metal termination is one of the focuses. The calculated results include the zero-temperature non-spin-polarized ground state energies, the non-uniform bond lengths, the band structures, the spatial charge densities, and the orbital-decomposed DOSs. The dependence on nanoribbon width, boundary structure and lithium/hydrogen passivation will be investigated thoroughly. Obviously, the non-uniform bond lengths could exist in GNR systems without/with lithium and hydrogen terminations, mainly owing to the transverse quantum-confinement effects. They strongly depend on the positions of carbon atoms, especially for those near the two edges. The predicted results could be verified by the high-resolution STM measurements. Energy bands might exhibit three kinds of dispersion relations, namely, the parabolic, linear and partially flat ones. Some of them are closely related to the edge atoms and hydrogens. The up-to-date angle-resolved photoemission spectroscopy (ARPES) examinations \cite{Rev.Mod.Phys.75(2003)473A.Damascelli, Phys.Scr.T109(2004)61A.Damascelli, Phys.Rev.Lett.81(1998)4943V.N.Strocov, Phys.Rev.Lett.79(1997)467V.N.Strocov} could account for part of theoretical predictions. The width-dependent energy gaps are calculated for various AGNRs in the absence and presence of hydrogen and alkali metal passivations. The band-edge states of band structure, which correspond to the van-Hove singularities \cite{Phys.Rev.89(1953)1189L.VanHove} in the wave-vector-energy space, display the special structures in DOS. In short, the chemical bondings associated with the spatial charge are proposed to explain the H-, alkali metal-, and edge-carbon-dominated energy bands as well as DOSs. The feature-rich GNRs are expected to have high potentials in near-future applications, such as, the electronic, spintronic, sensing, and energy devices.

\section{Geometry transition of alkalized and hydrogenated graphene nanoribbons}

\begin{figure*}
  \includegraphics[height=22cm, keepaspectratio]{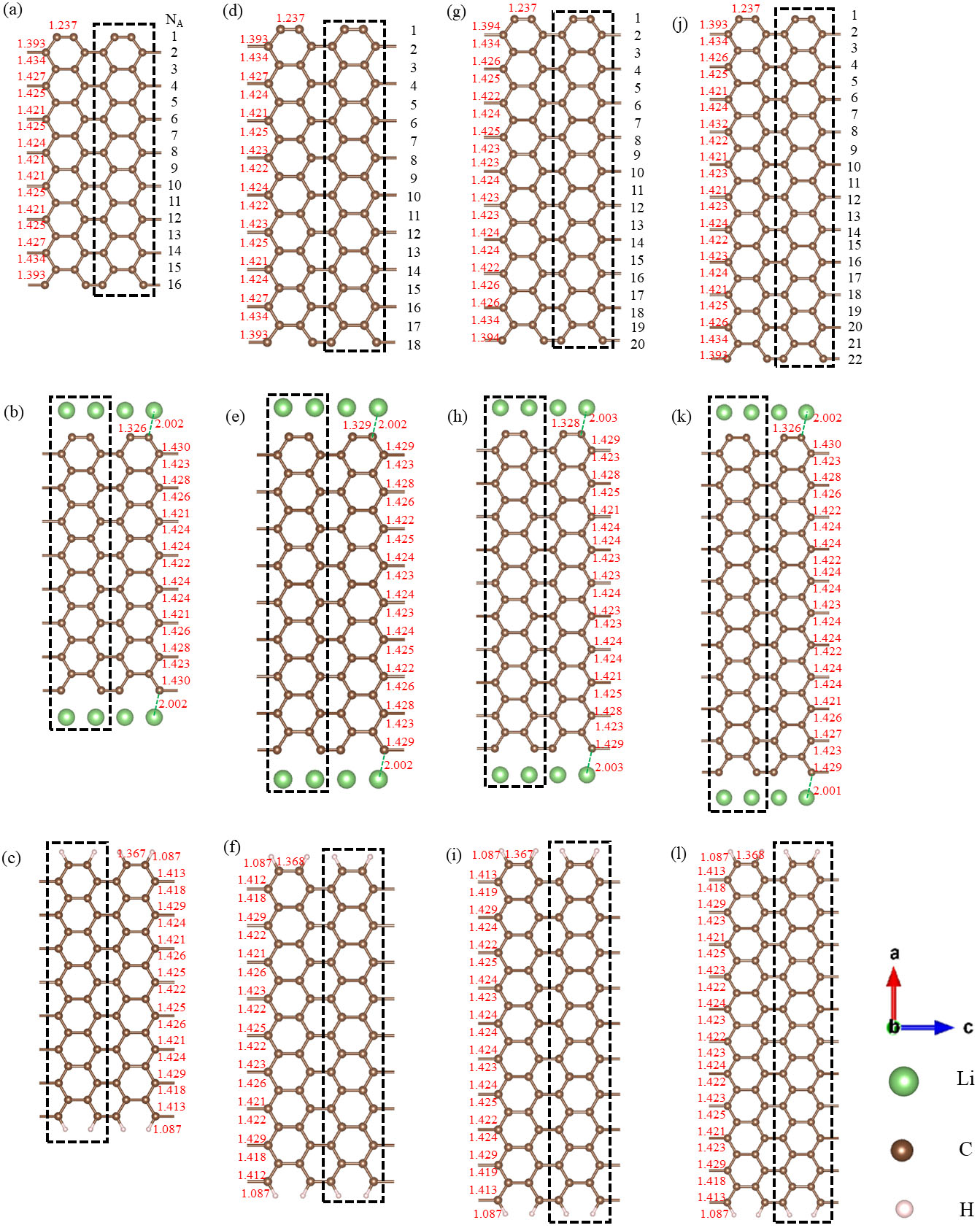}\\
  \caption{
(Color online)
The geometric structures of pristine (a)/(b)/(g)/(j), Li-passivated (b)/(e)/(h)/(k), and H-passivated (c)/(f)/(i)/(l) AGNR with $N_A=16$/18/20/22.
}
\label{fig:Figure01}
\end{figure*}

In this work, the geometric and electronic properties of pristine and hydrogen-/alkali metal-passivated AGNRs with various widths are investigated. The widths are characterized by the number of dimer lines $N_A$ along $\hat{a}$ axis. We perform four different widths of $N_A=16$, 18, 20, and 22 as shown in Fig.~\ref{fig:Figure01}.

The $N_A=n$ (where $n=16$, 18, 20, and 22) pristine AGNR contains $2n$ carbon atoms in a unit cell, and Li- and H-passivated ones possess $2n$ carbon atoms and 4 adatoms. The armchair boundaries are easy in creating the edge reconstruction. Elements, with an atomic number of less than 20, can serve as the edge-passivated adatoms to fully illustrate the decoration effects. An obvious charge transfer, being induced by the significant chemical bondings between the edge carbon atoms and the adatoms, is evaluated from the Bader analysis method. Moreover, the adsorption energy, the reduced total energy due to the adatoms bonding with the edge carbons, is characterized as the total energy difference between an edge-decorated GNR and two independent sub-systems (pristine nanoribbon and isolated adatoms). Specially, the former might have part of the total energy arising from a mechanical strain. In addition, the delicate calculations on the various total energies are done for the same unit cell.

For the pristine cases, since the ribbon edge is not saturated with hydrogen atoms, the geometric reconstruction takes place, leading to the edge dangling bonds with the shorter bond lengths. For example, the $N_A=16$ system has the outermost C--C length of 1.237 {\AA} (Fig.~\ref{fig:Figure01}(a)), in which the C--C bonds are non-uniform, and their central lengths are close to those of monolayer graphene. Such feature is hardly changed by the variation of ribbon width. The unstable dangling bonds provide suitable environment to form the C--H and C--Li bondings. On the other hand, the hydrogen-terminated and lithium-terminated cases exhibit the strong C--H $sp^2$ bonding with, respectively, 1.087 {\AA} (Fig.~\ref{fig:Figure01}(b)) and 2.002 {\AA} (Fig.~\ref{fig:Figure01}(c)) bond length, which are slightly affected by different widths. The edge C--C bond lengths are quite different between these three cases. The calculated results exhibit that the edge C--C bond lengths of Li- and H-passivated systems are, respectively, 1.326/1.329/1.328/1.326 {\AA} and 1.367/1.368/1.367/1.368 {\AA} for $N_A=16$/18/20/22, and all are longer than those of pristine systems of 1.237/1.237/1.237/1.237 {\AA}. Apparently, the boundary terminations have dramatically changed the bond lengths and thus the hopping integrals, especially for those near the two edges. However, they cannot destroy the $\pi$ and $\sigma$ bondings, respectively, due to the $2p_z$ and $(2p_x, 2p_y)$ orbitals; that is, they do not affect the orbital dominance for the low- and high-energy (deep-energy) electronic states.

According to the Bader methods, in the H-passivated systems, a tiny electric charge of 0.02--0.04 $e$ is transferred from the H atom towards the outermost C atom. (See Table~\ref{tab:table01}.) This reveals that the C--H bonds appear as a protruding edge structure, bit not one part of a polygon. This is to say, the H adatoms cannot bridge the strong interactions of carbon atoms on the neighboring edge dimers. Apparently, the $1s$ orbital is not sufficient in creating the multi-chemical bondings. The Li-passivated cases, however, lose more electron charge of 0.511--0.538 $e$ from Li atom to carbon atoms. On the other hand, each Li atom contributes more than half of the outmost electron charges to the neighboring carbon atoms. The high charge transfer indicates the highly ionic behavior of adatom.


\begin{table*}
\caption{
\label{tab:table01}
The edge passivation of $N_A=16$, 18, 20, and 22 AGNR.
}
\begin{ruledtabular}
\begin{tabular}{cccccc}
$N_A$ & Atom & Outmost C--C ({\AA}) & X--C ({\AA})\footnote{X can be Li or H.} & X--X ({\AA}) & Charge transfer of X ($e$)\footnote{The minus sign denotes electrons which are lost.} \\
\hline
   & C  & 1.237 &       &       &        \\
16 & Li & 1.326 & 2.002 & 2.139 & -0.538 \\
   & H  & 1.367 & 1.087 & 1.856 & -0.032 \\
\hline
   & C  & 1.237 &       &       &        \\
18 & Li & 1.329 & 2.002 & 2.138 & -0.518 \\
   & H  & 1.368 & 1.087 & 1.856 & -0.041 \\
\hline
   & C  & 1.237 &       &       &        \\
20 & Li & 1.328 & 2.003 & 2.139 & -0.511 \\
   & H  & 1.367 & 1.087 & 1.855 & -0.025 \\
\hline
   & C  & 1.237 &       &       &        \\
22 & Li & 1.326 & 2.002 & 2.140 & -0.521 \\
   & H  & 1.368 & 1.087 & 1.856 & -0.031
\end{tabular}
\end{ruledtabular}
\end{table*}

\section{Band-structure transformation of alkalized and hydrogenated graphene nanoribbons}

\subsection{Band structure}

\begin{figure*}
  \includegraphics[height=22cm, keepaspectratio]{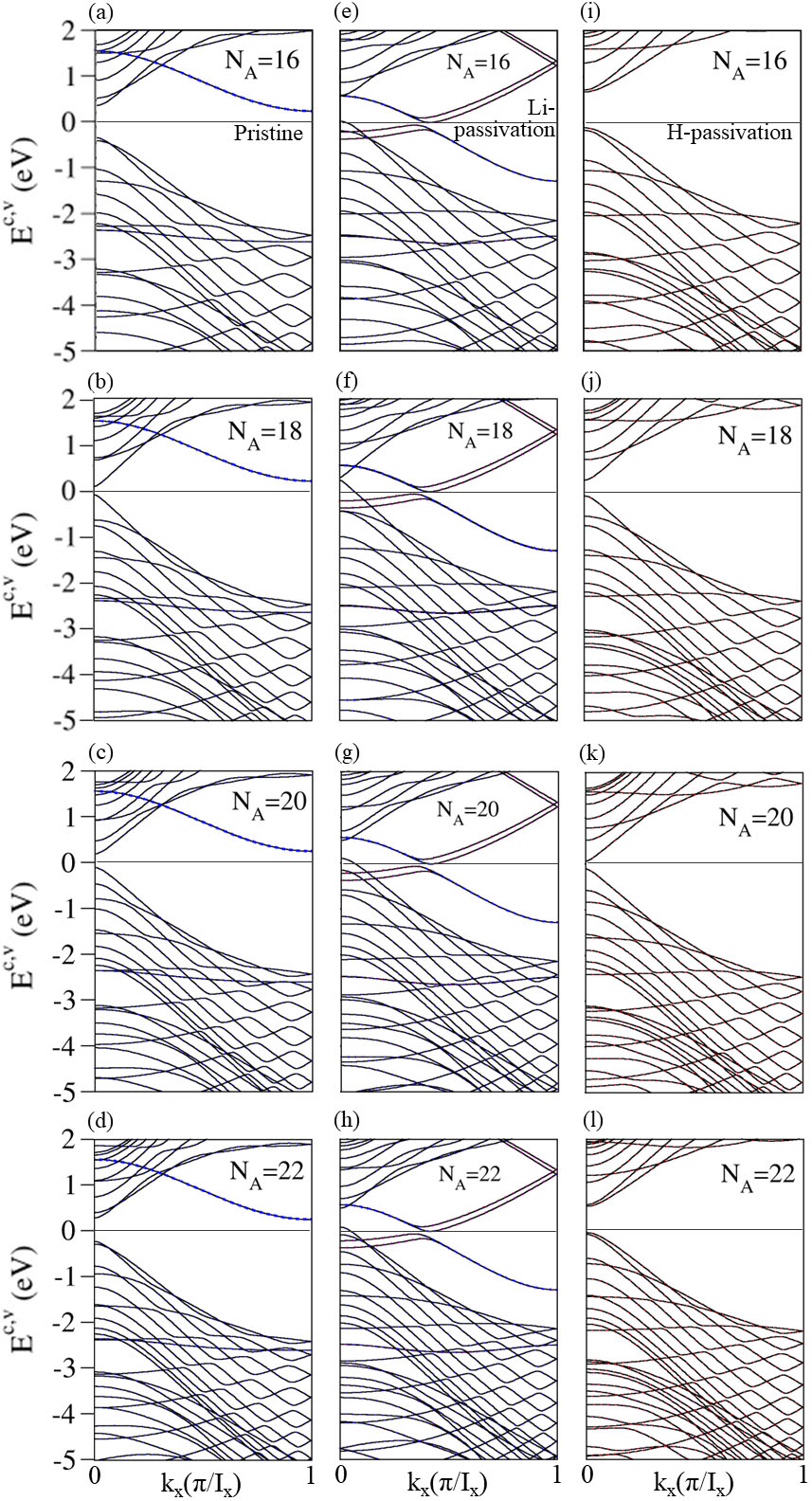}\\
  \caption{
(Color online)
The band structures of pristine (a)/(b)/(c)/(d), Li-passivated (e)/(f)/(g)/(h), and H- passivated (i)/(j)/(k)/(l) AGNR with $N_A=16$/18/20/22.
}
\label{fig:Figure02}
\end{figure*}

The strong orbital hybridizations in various chemical bonds, combined with the finite-size effects, can greatly diversify the electronic properties. A pristine $N_A=16$ AGNR exhibits an indirect band gap of $E_g=0.482$ eV. The highest occupied state (HOS) and the lowest unoccupied state (LUS) are situated at $k_x=0$ and 1, respectively. The former and the latter are, respectively, dominated by the non-edge and edge carbon atoms. Moreover, the dangling bonds at the edges are responsible for two pairs of valence and conduction bands. The main features of band structures, atom dominance, energy dispersion and band gap, are dramatically changed after the significant edge passivation, as clearly indicated in Fig.~\ref{fig:Figure02}. Apparently, the edge-carbon-dominated energy bands disappear in the presence of X--C bonds, i.e., the absence of dangling bonds will thoroughly modify the electronic structures. The low-lying energy bands depend on the kinds of adatoms.

Electronic properties of pristine AGNRs are mainly determined by the edge structure, dangling bonds, and quantum confinement effects. The unoccupied conduction bands are asymmetric to the occupied valence bands about the Fermi level ($E_F=0$), as shown in Fig.~\ref{fig:Figure02}(a) for the $N_A=16$ armchair system. Most energy bands exhibit strong parabolic dispersions that belong to the extended states with wide spatial distributions. There exist four energy bands with localized carrier distributions near the ribbon edges, in which the radius of each circle represents the contribution percentage from the edge atoms. They principally come from the edge dangling bonds. These edge-atom-dominated conduction and valence bands are doubly degenerate near $k_x=1$; furthermore, the valence ones possess weak energy dispersions at about $-2.5~\mathrm{eV} \leq E^v \leq -2.2~\mathrm{eV}$. The ARPES measurements on them are absent up to now. They might be associated with specific energy gaps. the edge-atom-induced conduction ones at $k_x=1$ and the first valence band (counted from $E_F$) at $k_x=0$, respectively, correspond to the LUS and the HOS. These two kinds of armchair systems belong to indirect-gap semiconductors, e.g., $N_A=16$ and 22 in Figs.~\ref{fig:Figure02}(a) and \ref{fig:Figure02}(d), respectively. On the other hand, the $N_A=18$ and 20 systems, as indicated in Figs.~\ref{fig:Figure02}(b) and \ref{fig:Figure02}(c) respectively, possess the LUS and HOS in the first pair of conduction and valence bands at $k_x=0$; that is, they are direct-gap semiconductors. Specifically, indirect and direct energy gaps could be examined from the experimental measurements of optical spectra and transport properties.

The lithium and hydrogen edge terminations dramatically alter the electronic properties of AGNRs, as shown in Figs.~\ref{fig:Figure02}(e) to \ref{fig:Figure02}(l). As for H-passivated systems, the edge-atom-created conduction bands and weakly dispersive valence band vanish for various H-passivated armchair systems, indicating the absence of dangling bonds. The strong C--H bondings and the very low bound state energy of H-$1s$ orbital are responsible for the drastic changes in energy bands. The edge C and H atoms make important contributions to certain valence bands in the range of $-6~\mathrm{eV} \leq E^v \leq -3~\mathrm{eV}$. The C-H-dominated energy band width is wider than 2.5 eV. This further requires the high-resolution ARPES verifications. The orbital hybridizations in C--C bonds will be analyzed using the orbital-decomposed DOSs. The LUS and HOS, respectively, correspond to the first conduction and the first valence band (counted from $E_F$) at $k_x=0$. Hence, all the H-passivated AGNRs belong to semiconductors with direct band gaps. As for Li-passivated systems, the gap vanishes after Li-passivation. The lowest conduction band and the highest valence band slightly intersect the Fermi level, respectively, at $k_x=0.46$ and 0. This indicates the Li-passivated systems perform as a semi-metal with co-existed electrons and holes, as shown in Figs.~\ref{fig:Figure02}(e) to \ref{fig:Figure02}(h). In general, the predicted energy gaps could be examined by the experimental measurements of band structure, DOS, optical spectra, and transport conductance.

\subsection{Spatial charge}

\begin{figure*}
  \includegraphics[keepaspectratio]{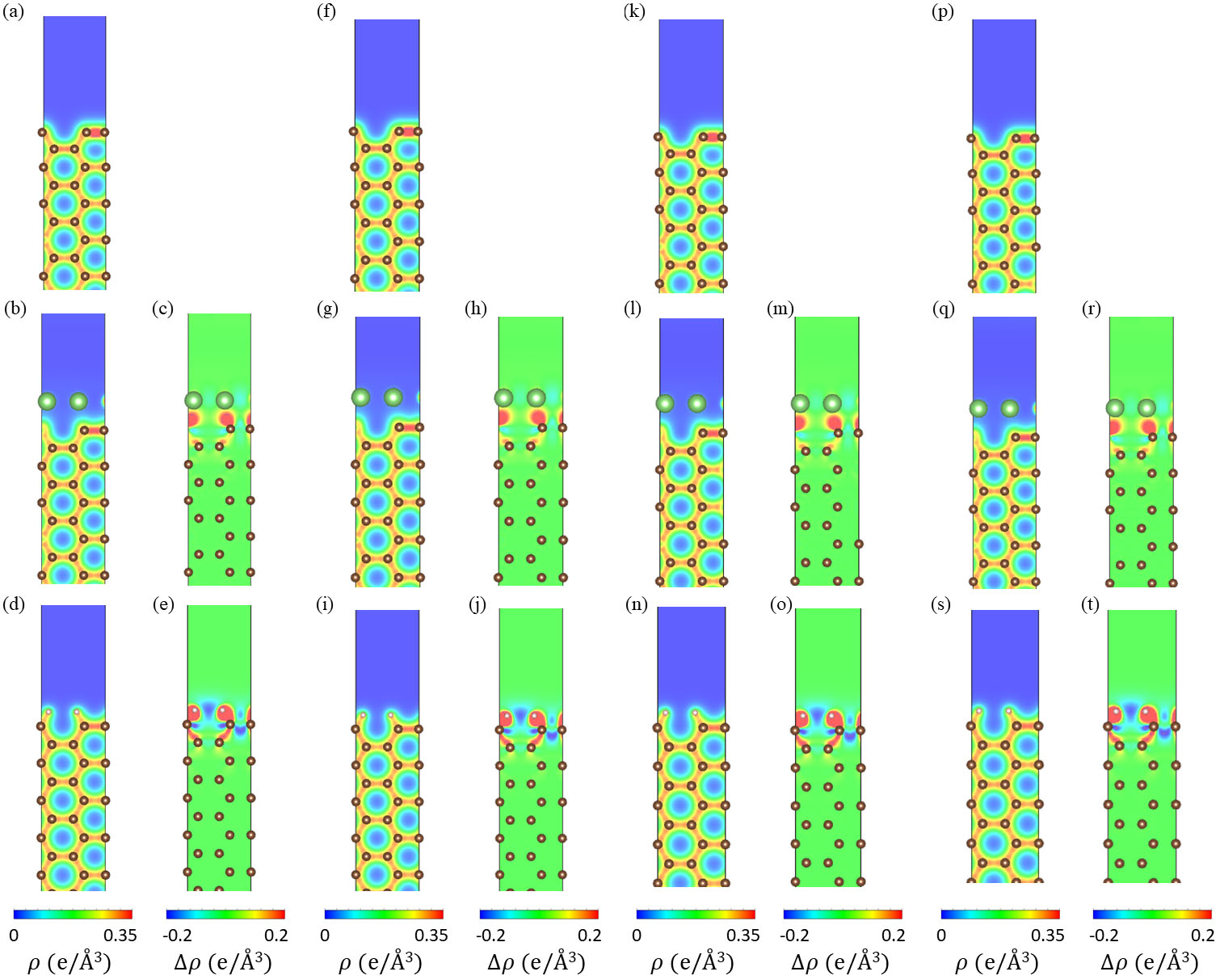}\\
  \caption{
(Color online)
The spatial charge distributions of pristine (a)/(f)/(k)/(p), Li-passivated (b)/(g)/(l)/(q), and H-passivated (d)/(i)/(n)/(s) AGNR; and the charge differences of Li-passivated (c)/(h)/(m)/(r), and H-passivated (e)/(j)/(o)/(t) AGNR with $N_A=16$/18/20/22.
}
\label{fig:Figure03}
\end{figure*}

The carrier density ($\rho$) and the variation of carrier density ($\Delta \rho$) can provide useful information on the orbital bondings and energy bands. The former directly reveals the bonding strength of C--C bonds near the center or edge of graphene nanoribbon, as illustrated in Fig.~\ref{fig:Figure03}. All the C--C bonds possess the strong covalent $\sigma$ bonds and the relatively weak $\pi$ bonds simultaneously. Such bondings are enhanced for the edge C--C bonds in pristine systems, as shown by the dark red region. This is responsible for the shortened bond lengths and the energy range of edge-atom-dominated bands. As for H-terminated ones, the C--H bondings are revealed by the red region between them, while the edge C--C bonds are weakened under the loss of the $\sigma$ electrons in Figs.~\ref{fig:Figure03}(d), \ref{fig:Figure03}(i), \ref{fig:Figure03}(n), and \ref{fig:Figure03}(s). As for Li-passivated ones, the edge C--C bonds weaken slightly as well. These features reflect the variation in bond lengths and can be further analyzed by the variation of carrier density. $\Delta \rho$ is obtained by subtracting the carrier density of an isolated carbon from that of an AGNR. Carbon atoms contribute four valence electrons to create two specific chemical bondings, namely the strong $\sigma$ bonding of $(2s, 2p_x, 2p_y)$ orbitals and the $\pi$ bonding of $2p_z$ orbitals. The former induces the increased charge density between two carbon atoms. There are more charges distributed at the ribbon edge (dark red), indicating the strength of the C--C triple bond. $\Delta \rho$ in H-passivated AGNRs presents a clear image of charge transfer from edge carbons to hydrogen atoms or lithium atoms. This accounts for the vanishing edge-atom-created energy bands. We can find that the $\Delta \rho$ between C--C bonds is almost zero near the center of ribbons whether with or without passivation.

\subsection{DOS}

\begin{figure*}
  \includegraphics[keepaspectratio]{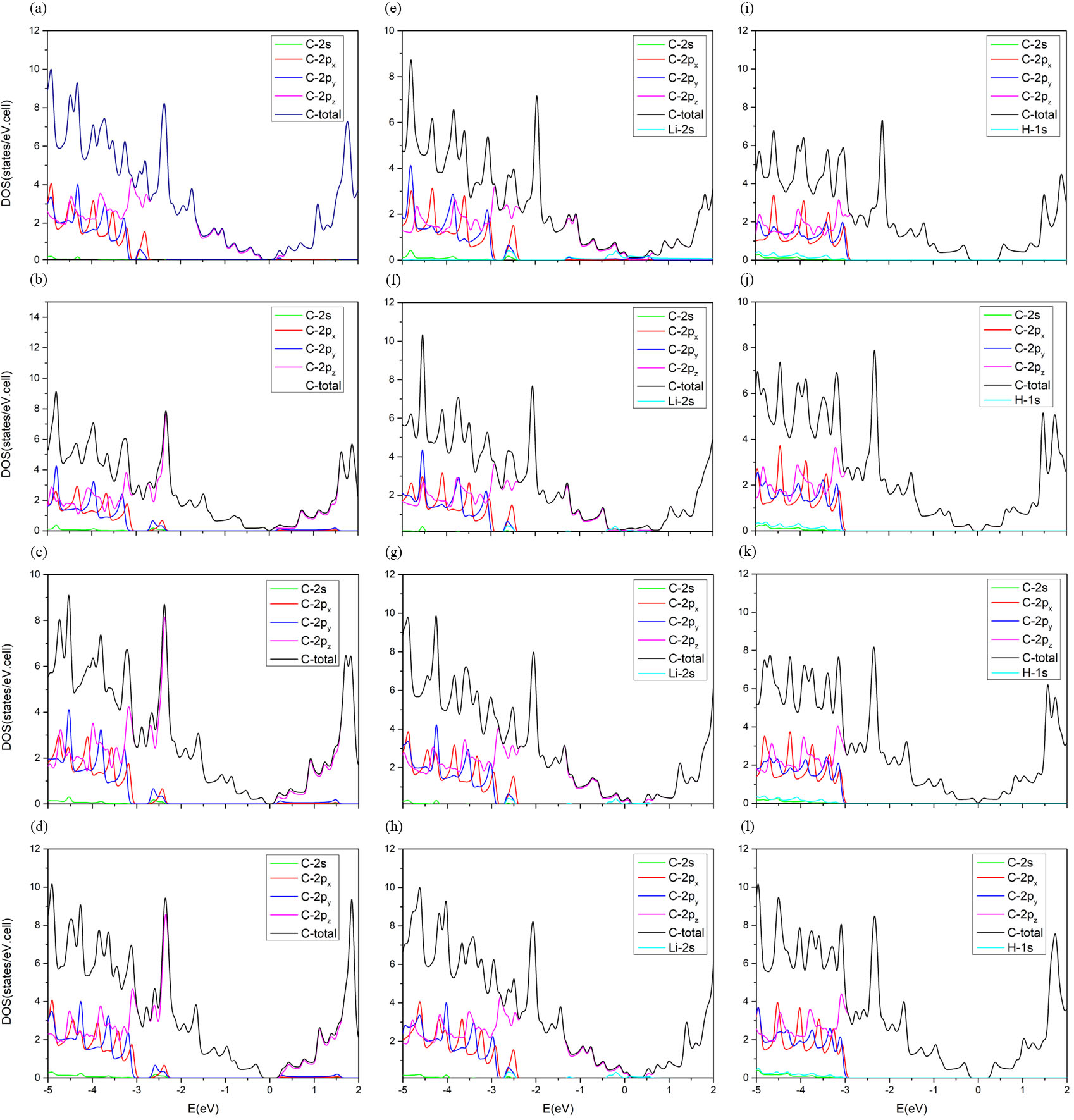}\\
  \caption{
(Color online)
The density of states of pristine (a)/(b)/(c)/(d), Li-passivated (e)/(f)/(g)/(h), and H-passivated (i)/(j)/(k)/(l) AGNR with $N_A=16$/18/20/22.
}
\label{fig:Figure04}
\end{figure*}

DOSs directly reflect the main features of band structures, in which the orbital-decomposed ones can understand the orbital hybridizations in C--C and C--H bonds. There are a lot of special structures in DOSs due to the band-edge states. Most van Hove singularities, which come from parabolic energy dispersions, appear in the square-root-divergent form because of the 1D characteristics, as those indicated in Fig.~\ref{fig:Figure04}. Their heights are inversely proportional to the band curvatures, The number of asymmetric prominent peaks increases as the nanoribbon width grows. A pair of asymmetric peaks, which is centered at the Fermi level, could characterize the energy gap. Such structures might not available in distinguishing the direct- or indirect-gap characteristics, since the low-lying double peaks are almost absent in the conduction band owing to the broadening effect. That is to say, the scanning tunneling spectroscopy (STS) measurements could not be utilized to identify the direct or indirect gaps. Specifically, the edge carbon atoms make important contributions to the first conduction-band peak nearest to $E_F$. Few special structures exhibit in the delta-function-like form corresponding to the partially flat edge-atom-dominated valence energy bands in pristine armchair systems, e.g., DOSs in the range of $-2.5~\mathrm{eV} \leq E^v \leq -2.2~\mathrm{eV}$ in Figs.~\ref{fig:Figure04}(a) to \ref{fig:Figure04}(d). They become asymmetric prominent peaks at deeper energies, as indicated in Figs.~\ref{fig:Figure04}(i) to \ref{fig:Figure04}(l) below $E < -3$ eV by the light-blue peaks and those accompanied with them for H-passivated systems. The density of states for Li-passivated systems, however, exhibits a non-zero value around the Fermi level ($E_F=0$), which reveals the co-dominated Li-$1s$ and C-$2p_z$ orbitals and agrees well with the energy band structures in Figs.~\ref{fig:Figure04}(e) to \ref{fig:Figure04}(h).

The orbital-projected DOSs show that the main contributions are due to the outer four orbitals of the carbon atom, $1s$ orbital of the hydrogen atom, and $2s$ orbital of the lithium atom. For pristine armchair systems, the $\pi$ and $\sigma$ pronounced peaks, respectively, occur at the valence range of $E < 0$ and $E < -3$ eV, while both of them are revealed at the whole conduction range, mainly owing to the edge carbons with the dangling bonds. This indicates the existence of the partial $sp^3$ near the armchair boundaries. The low-lying conduction structures are only dominated by the $2p_z$ orbitals after the hydrogen terminations; that is, the $\pi$ bonding determines the low-energy essential properties of the Li-passivated and H-passivated AGNRs. Furthermore, the $1s$- and $(2p_x, 2p_y)$-related asymmetric peaks happen simultaneously in H-passivated systems, revealing the strong orbital hybridizations among three orbitals. There exist the $sp^2$ chemical bonding in the C--H bonds. As for the Li-passivated ones, the $2s$ orbital of Li atom co-dominate with $(2p_x, 2p_y)$ orbitals of C atom at the range of $-2.8~\mathrm{eV} \leq E \leq -2.4~\mathrm{eV}$, and with $2p_z$ orbital of C atom at the range near Fermi level ($E_F=0$). This indicates the existence of the partial $sp^3$ near the boundaries similar to the pristine ones.

\section{Concluding remarks}

In this work, we have demonstrated theoretical studies on pristine GNRs, even with/without hydrogen atom and alkali metal atom passivation. The electronic properties are sensitive to the change in edge structure. On the other hand, the band gaps are predicted to be inversely related to the ribbon width. The calculated results contain the zero-temperature non-spin-polarized ground state energies, the non-uniform bond lengths, the band structures, the spatial charge densities, and the orbital-decomposed DOSs. Apparently, the non-uniform bond lengths can exist in GNR systems even without lithium and hydrogen terminations, mainly due to the transverse quantum-confinement effects. They strongly depend on the positions of carbon atoms, especially for those near the two edges. The width-dependent energy gaps are calculated for various AGNRs in the absence and presence of hydrogen and alkali metal passivations. The band-edge states corresponding to the van-Hove singularities in the wave-vector-energy space exhibit the special structures in DOS. In short, the chemical bondings associated with the spatial charge are proposed to explain the H-, alkali metal-, and edge-carbon-dominated energy bands as well as DOSs.

Up to now, there are many ways to fabricate GNRs. The main obstacles and disadvantages leading GNRs from real applications are the width and edge modulations, the substrate effect, edge termination, and defects. Electronic properties, especially the energy gap, are sensitive to the width and edge structure of nanoribbons. They are also altered by charge transfer between the substrate and GNRs. During some fabrication processes, the ribbon edges might be partly terminated with hydrogen, oxygen, and other functional groups; therefore, the edge-related properties will be changed. Non-hexagonal defects (often pentagonal and heptagonal structures) on the ribbon plane cause drastic changes in electronic properties. Some of the passivation effects are discussed in this work. To date, the researchers and engineers are continuously finding other routes of high yield to precisely control the nanoscale width and perfect edge structure and trying to overcome the technical problems for real applications.

The commercialization and popularization of electric vehicles (EVs) worldwide \cite{engrxiv2020OutlooksLiIonBatteriesH.C.Chung, BookChChung2021EngIntePotentialAppOutlooksLiIonBatteryIndustry, GlobalEVOutlook2020IEA} lead to the widespread and mass production of lithium-ion (Li-ion) batteries. The retired power batteries of EVs and buses have largely increased, leading to resource waste and environmental protection threats. Hence, recycling and utilization of such retired batteries have been promoted \cite{Sustain.EnergyTechnol.Assess.6(2014)64L.Ahmadi, Renew.Sustain.EnergyRev.93(2018)701E.Martinez-Laserna, CellRep.Phys.Sci.2(2021)100537J.Zhu}. Some retired power batteries still possess about 80\% initial capacity \cite{J.Environ.Manage.232(2019)354L.C.Casals, FMEAofLFPBatteryModule2018Chung, WorldElectr.Veh.J.9(2018)24A.Podias, J.EnergyStorage11(2017)200S.Tong, J.PowerSources196(2011)5147E.Wood}. Hence, they can be repurposed and utilized once again, e.g., serving as the battery modules in the stationary energy storage system \cite{Batteries3(2017)10L.C.Casals, EnergyPolicy71(2014)22C.Heymans, WasteManage.113(2020)497D.Kamath, Batteries5(2019)33H.Quinard}. Governments in various countries have noticed this serious issue and prepared to launch their policies to deal with the recovery and reuse of repurposing batteries, such as coding principles, traceability management systems, manufacturing factory guidelines, dismantling process guidelines, residual energy measurement, federal and state tax credits, rebates, and other financial support  \cite{Sci.Data8(2021)165H.C.Chung, J.TaiwanEnergy6(2019)425H.C.Chung, MJTE860(2020)35H.C.Chung, BookChChung2021UL1974, EnergyPolicy113(2018)535K.Gur, IEEEAccess7(2019)73215E.Hossain}. From the discovery of graphene and more low-dimensional materials \cite{Sci.Rep.9(2019)2332H.C.Chung, BookLinFirstPrinciplesCathodeElectrolyteAnodeBatteryMaterials}, scientists believe that these materials (such as 1D finite-width graphene nanoribbon systems discussed in this work) can serve as potential additives in battery Materials (such as cathode, anode, and electrolyte) to improve the performance and recycling problems of Li-ion batteries.

\begin{acknowledgments}
The author (H. C. Chung) thanks Prof. Ming-Fa Lin for the book chapter invitation and for inspiring him to study this topic. H. C. Chung would like to thank the contributors to this article for their valuable discussions and recommendations, Jung-Feng Jack Lin, Hsiao-Wen Yang, Yen-Kai Lo, and An-De Andrew Chung. The author (H. C. Chung) thanks Pei-Ju Chien for English discussions and corrections as well as Ming-Hui Chung, Su-Ming Chen, Lien-Kuei Chien, and Mi-Lee Kao for financial support. This work was supported in part by Super Double Power Technology Co., Ltd., Taiwan, under the project “Development of Cloud-native Energy Management Systems for Medium-scale Energy Storage Systems (\href{https://osf.io/7fr9z/}{https://osf.io/7fr9z/})” (Grant number: SDP-RD-PROJ-001-2020). This work was supported in part by Ministry of Science and Technology (MOST), Taiwan (grant numbers: MOST 108-2112-M-006-016-MY3, MOST 109-2124-M-006-001, and MOST 110-2811-M-006-543).
\end{acknowledgments}

\bibliography{Reference}
\bibliographystyle{apsrev4-2}

\end{document}